\documentclass[10pt,conference]{IEEEtran}
\usepackage{fixltx2e}
\usepackage{cite}
\usepackage{url}
\usepackage{mathrsfs}
\usepackage{color}
\usepackage{float}
\usepackage[caption=false]{subfig}

\ifCLASSINFOpdf
   \usepackage[pdftex]{graphicx}
   \graphicspath{{Figs/}}
   \DeclareGraphicsExtensions{.pdf,.jpeg,.png}
\else
\fi

\usepackage[cmex10]{amsmath}
\usepackage{amsmath}
\usepackage{amssymb}
\usepackage{amsthm}
\usepackage{amsfonts}
\usepackage{bm}
\usepackage{xfrac}
\usepackage{empheq}
\usepackage[normalem]{ulem} 
\usepackage{soul} 
\usepackage{mathtools}

\newtheorem{theorem}{Theorem}

\newtheorem{lemma}{Lemma}
\newtheorem*{observ}{Observation}

\newtheorem{corollary}{Corollary}
\newtheorem{remark}{Remark}
\theoremstyle{definition}

\usepackage[a4paper,bindingoffset=0.2in,%
left=1in,right=1in,top=1in,bottom=1in,%
footskip=.25in]{geometry}
\graphicspath{{figs/}}

\begin{document}
	\newgeometry{left=0.7in,right=0.7in,top=.5in,bottom=1in}
	\title{New Privacy Mechanism Design With Direct Access to the Private Data}
\vspace{-5mm}
\author{
		\IEEEauthorblockN{Amirreza Zamani, Tobias J. Oechtering, Mikael Skoglund \vspace*{0.5em}
			\IEEEauthorblockA{\\
                              Division of Information Science and Engineering, KTH Royal Institute of Technology \\
				Email: \protect amizam@kth.se, oech@kth.se, skoglund@kth.se }}
		}
	\maketitle

\begin{abstract}
	The design of a statistical signal processing privacy problem is studied where the private data is assumed to be observable. 
	In this work, 
	an agent observes useful data $Y$, which is correlated with private data $X$, and wants to disclose the useful information to a user. A statistical privacy mechanism is employed to generate data $U$ based on $(X,Y)$ that maximizes the revealed information about $Y$ while satisfying a privacy criterion. 
	 
	To this end, we use extended versions of the Functional Representation Lemma and Strong Functional Representation Lemma and combine them with a simple observation which we call separation technique. 
	New lower bounds on privacy-utility trade-off are derived and we show that they can improve the previous bounds. We study the obtained bounds in different scenarios and compare them with previous results. 
\end{abstract}
\section{Introduction}

\begin{figure}[]
	\centering
	\includegraphics[scale = .15]{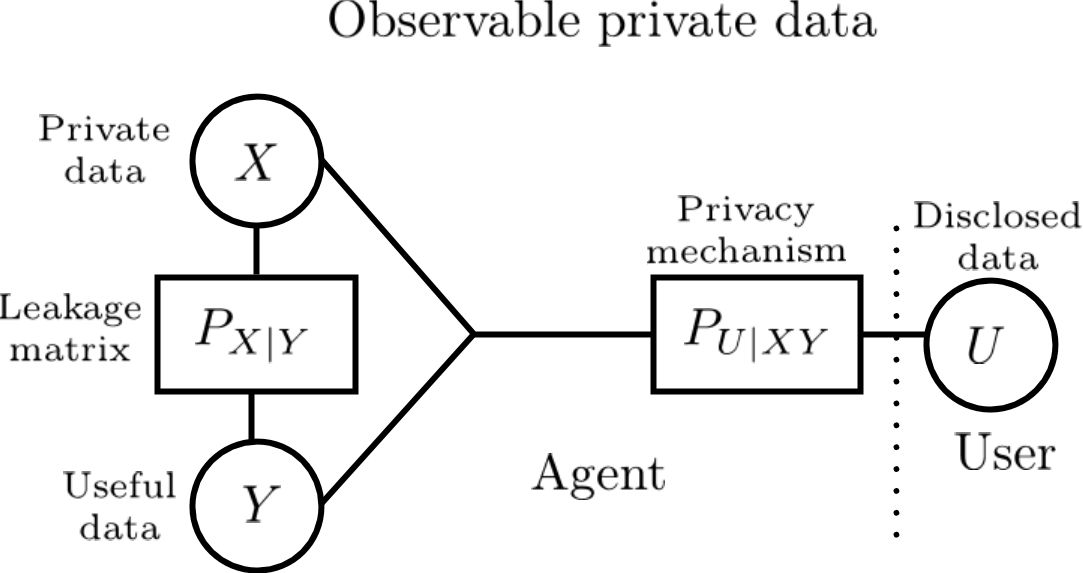}
	\caption{Considered scenario where the agent has access to both $X$ and $Y$.}
	\label{ISITsys}
\end{figure}

In this paper, random variable (RV) $Y$ denotes the useful data and is correlated with the private data denoted by RV $X$. Furthermore, disclosed data is described by RV $U$. In this work, an agent wants to disclose the useful information to a user as shown in Fig.~\ref{ISITsys}. The agent has direct access to both $X$ and $Y$, i.e., the agent observes $(X,Y)$. The goal is to design $U$ based on $(X,Y)$ that reveals as much information as possible about $Y$ and satisfies a privacy criterion. We use mutual information to measure utility and privacy leakage. In this work, some bounded privacy leakage is allowed, i.e., $I(X;U)\leq \epsilon$. 
\\
The privacy mechanism design problem is receiving increased attention in information theory recently. Related works can be found in 
\cite{Calmon2,yamamoto, sankar,borz, gun,khodam,Khodam22,kostala,king1,issa, makhdoumi, dwork1, calmon4, issajoon, asoo, Total,deniz3, issa2}. 
In \cite{Calmon2}, fundamental limits of the privacy utility trade-off measuring the leakage using estimation-theoretic guarantees are studied.
In \cite{yamamoto}, a source coding problem with secrecy is studied.
Privacy-utility trade-offs considering equivocation as measure of privacy and expected distortion as a measure of utility are studied in both \cite{yamamoto} and \cite{sankar}.
In \cite{borz}, the problem of privacy-utility trade-off considering mutual information both as measures of privacy and utility given the Markov chain $X-Y-U$ is studied. It is shown that under perfect privacy assumption, i.e., $\epsilon=0$, the privacy mechanism design problem can be reduced to a linear program. This work has been extended in \cite{gun} considering the privacy utility trade-off with a rate constraint for the disclosed data.
Moreover, in \cite{borz}, it has been shown that information can be only revealed if $P_{X|Y}$ is not invertible. In \cite{khodam}, we designed privacy mechanisms with a per letter privacy criterion considering an invertible $P_{X|Y}$ where a small leakage is allowed. We generalized this result to a non-invertible leakage matrix in \cite{Khodam22}.

Here we consider the problem studied in  \cite{borz}, \cite{kostala}, \cite{shahab}, and \cite{king1}. In \cite{kostala}, the problem of \emph{secrecy by design} is studied where the results are derived under the perfect secrecy assumption, i.e., no leakages are allowed which corresponds to $\epsilon=0$. Bounds on secure decomposition have been derived using the Functional Representation Lemma and new bounds on privacy-utility trade-off are derived. The bounds are tight when the private data is a deterministic function of the useful data. In \cite{king1}, the privacy problems considered in \cite{kostala} are generalized by relaxing the perfect secrecy constraint and allowing some leakages. More specifically, we considered bounded mutual information, i.e., $I(U;X)\leq \epsilon$ for privacy leakage constraint.
Furthermore, in the special case of perfect privacy we derived a new upper bound for the perfect privacy function and it has been shown that this new bound generalizes the bound in \cite{kostala}. Moreover, it has been shown that the bound is tight when $|\mathcal{X}|=2$.

In the present work, we generalize the lower bounds obtained in \cite{king1}. To this end, we use extended versions of the Functional Representation Lemma and the Strong Functional Representation Lemma and combine them with a simple observation. The Functional Representation Lemma and the Strong Functional Representation Lemma are constructive lemmas that are valuable for the privacy design.
The simple observation corresponds to representing a discrete random variable (RV) by two RVs which are correlated in general. We call the observation \emph{separation technique} since it separates a RV into two RVs. We show that new lower bounds improve the bounds obtained in \cite{king1}. In Section \ref{khaye}, we study the bounds in special cases and compare them with \cite{king1}.   

\section{system model and Problem Formulation} \label{sec:system}
Let $P_{XY}$ denote the joint distribution of discrete random variables $X$ and $Y$ defined on alphabets $\cal{X}$ and $\cal{Y}$. We assume that cardinality $|\mathcal{X}|$ is finite and $|\mathcal{Y}|$ is finite or countably infinite.
We represent $P_{XY}$ by a matrix defined on $\mathbb{R}^{|\mathcal{X}|\times|\mathcal{Y}|}$ and 
marginal distributions of $X$ and $Y$ by vectors $P_X$ and $P_Y$ defined on $\mathbb{R}^{|\mathcal{X}|}$ and $\mathbb{R}^{|\mathcal{Y}|}$ given by the row and column sums of $P_{XY}$. 
We represent the leakage matrix $P_{X|Y}$ by a matrix defined on $\mathbb{R}^{|\mathcal{X}|\times|\cal{Y}|}$.

Here, we use mutual information as utility and leakage measures. The privacy mechanism design problem can be stated as follows 
\begin{align}
h_{\epsilon}(P_{XY})&=\sup_{\begin{array}{c} 
	\substack{P_{U|Y,X}: I(U;X)\leq\epsilon,}
	\end{array}}I(Y;U).\label{main1}
\end{align} 
The relation between $U$ and the pair $(Y,X)$ is described by the kernel $P_{U|Y,X}$ defined on $\mathbb{R}^{|\mathcal{U}|\times|\mathcal{Y}|\times|\mathcal{X}|}$.
 In the following we study the case where $0\leq\epsilon< I(X;Y)$, otherwise the optimal solution of $h_{\epsilon}(P_{XY})$ is $H(Y)$ achieved by $U=Y$. 
 \begin{remark}
 	\normalfont
 	For $\epsilon=0$, \eqref{main1} leads to the secret-dependent perfect privacy function $h_0(P_{XY})$, studied in \cite{kostala}, where upper and lower bounds on $h_0(P_{XY})$ have been derived. The bounds are generalized in \cite{king1}.
 \end{remark}

 \section{Main Results}\label{sec:resul}
 In this section, we first recall the Functional Representation Lemma (FRL) \cite[Lemma~1]{kostala}, Strong Functional Representation Lemma (SFRL) \cite[Theorem~1]{kosnane}, Extended Functional Representation Lemma (EFRL) \cite[Lemma 3]{king1}, and Extended Strong Functional Representation Lemma (ESFRL) \cite[Lemma 4]{king1}, for discrete $X$ and $Y$. We then present a simple but important result regarding representing a RV by two correlated RVs. 
 \begin{lemma}\label{lemma1} (Functional Representation Lemma \cite[Lemma~1]{kostala}):
 	For any pair of RVs $(X,Y)$ distributed according to $P_{XY}$ supported on alphabets $\mathcal{X}$ and $\mathcal{Y}$ where $|\mathcal{X}|$ is finite and $|\mathcal{Y}|$ is finite or countably infinite, there exists a RV $U$ supported on $\mathcal{U}$ such that $X$ and $U$ are independent, i.e., we have
 	\begin{align}\label{c1}
 	I(U;X)=0,
 	\end{align}
 	$Y$ is a deterministic function of $(U,X)$, i.e., we have
 	\begin{align}
 	H(Y|U,X)=0,\label{c2}
 	\end{align}
 	and 
 	\begin{align}
 	|\mathcal{U}|\leq |\mathcal{X}|(|\mathcal{Y}|-1)+1.\label{c3}
 	\end{align}
 \end{lemma}
 \begin{lemma}\label{lemma2} (Strong Functional Representation Lemma \cite[Theorem~1]{kosnane}):
 	For any pair of RVs $(X,Y)$ distributed according to $P_{XY}$ supported on alphabets $\mathcal{X}$ and $\mathcal{Y}$ where $|\mathcal{X}|$ is finite and $|\mathcal{Y}|$ is finite or countably infinite with $I(X,Y)< \infty$, there exists a RV $U$ supported on $\mathcal{U}$ such that $X$ and $U$ are independent, i.e., we have
 	\begin{align*}
 	I(U;X)=0,
 	\end{align*}
 	$Y$ is a deterministic function of $(U,X)$, i.e., we have 
 	\begin{align*}
 	H(Y|U,X)=0,
 	\end{align*}
 	$I(X;U|Y)$ can be upper bounded as follows
 	\begin{align*}
 	I(X;U|Y)\leq \log(I(X;Y)+1)+4,
 	\end{align*}
 	and 
 	$
 	|\mathcal{U}|\leq |\mathcal{X}|(|\mathcal{Y}|-1)+2.
 	$
 \end{lemma}
 \begin{lemma}\label{lemma3} (Extended Functional Representation Lemma \cite[Lemma 3]{king1}):
 	For any $0\leq\epsilon< I(X;Y)$ and pair of RVs $(X,Y)$ distributed according to $P_{XY}$ supported on alphabets $\mathcal{X}$ and $\mathcal{Y}$ where $|\mathcal{X}|$ is finite and $|\mathcal{Y}|$ is finite or countably infinite, there exists a RV $U$ supported on $\mathcal{U}$ such that the leakage between $X$ and $U$ is equal to $\epsilon$, i.e., we have
 	\begin{align*}
 	I(U;X)= \epsilon,
 	\end{align*}
 	$Y$ is a deterministic function of $(U,X)$, i.e., we have  
 	\begin{align*}
 	H(Y|U,X)=0,
 	\end{align*}
 	and 
 	$
 	|\mathcal{U}|\leq \left[|\mathcal{X}|(|\mathcal{Y}|-1)+1\right]\left[|\mathcal{X}|+1\right].
 	$
 \end{lemma}
 \begin{proof}
 	The proof is based on adding a randomized response to the output of the FRL. The randomized response has been introduced in \cite{warner1965randomized}.
 \end{proof}
 \begin{lemma}\label{lemma4} (Extended Strong Functional Representation Lemma \cite[Lemma 4]{king1}):
 	For any $0\leq\epsilon< I(X;Y)$ and pair of RVs $(X,Y)$ distributed according to $P_{XY}$ supported on alphabets $\mathcal{X}$ and $\mathcal{Y}$ where $|\mathcal{X}|$ is finite and $|\mathcal{Y}|$ is finite or countably infinite with $I(X,Y)< \infty$, there exists a RV $U$ supported on $\mathcal{U}$ such that the leakage between $X$ and $U$ is equal to $\epsilon$, i.e., we have
 	\begin{align*}
 	I(U;X)= \epsilon,
 	\end{align*}
 	$Y$ is a deterministic function of $(U,X)$, i.e., we have 
 	\begin{align*}
 	H(Y|U,X)=0,
 	\end{align*}
 	$I(X;U|Y)$ can be  upper bounded as follows 
 	\begin{align*}
 	I(X;U|Y)\leq \alpha H(X|Y)+(1-\alpha)\left[ \log(I(X;Y)+1)+4\right],
 	\end{align*}
 	and 
 	$
 	|\mathcal{U}|\leq \left[|\mathcal{X}|(|\mathcal{Y}|-1)+2\right]\left[|\mathcal{X}|+1\right],
 	$
 	where $\alpha =\frac{\epsilon}{H(X)}$.
 \end{lemma}
 \begin{proof}
 	Similar to EFRL, the proof is based on adding a randomized response to the output of the SFRL.
 \end{proof}
 Next, we present a simple observation which we call ``separation technique''.
 \begin{observ}(Separation technique)
 	Any discrete RV $X$ supported on $\mathcal{X}=\{1,\ldots,|\mathcal{X}|\}$ can be represented by two RVs $(X_1,X_2)$. 
 \end{observ}
\begin{proof}
	First, let $|\mathcal{X}|$ be not a prime number. Thus, there exist $|\mathcal{X}_1|$ and $|\mathcal{X}_2|$ such that $|\mathcal{X}|=|\mathcal{X}_1|\times|\mathcal{X}_2|$ where $|\mathcal{X}_1|\geq|\mathcal{X}_2|\geq 2$. We can uniquely map each $x\in\mathcal{X}$ into a pair $(x_1,x_2)$ where $x_1\in\mathcal{X}_1$ and $x_2\in\mathcal{X}_2$. As a result, we can represent $X$ by the pair $(X_1,X_2)$ where $\mathcal{X}_1= \{1,\ldots,|\mathcal{X}_1|\}$, $\mathcal{X}_2= \{1,\ldots,|\mathcal{X}_2|\}$, and $P_{X}(x)=P_{X_1X_2}(x_1,x_2)$. Next, let $|\mathcal{X}|$ be a prime number. Hence, there exist $|\mathcal{X}_1|$ and $|\mathcal{X}_2|$ such that $|\mathcal{X}|+1=|\mathcal{X}_1|\times|\mathcal{X}_2|$ and we can represent $X$ by the pair $(X_1,X_2)$ where $P_{X_1X_2}(x_1=|\mathcal{X}_1|,x_2=|\mathcal{X}_2|)=0$. In other words, the last pair $(|\mathcal{X}_1|,|\mathcal{X}_2|)$ is not mapped to any $x\in\mathcal{X}$.   
\end{proof}
\begin{remark}
	The representation obtained by the separation technique is not unique. For instance, let $|\mathcal{X}|=16$. In this case, $|\mathcal{X}_1|=4$, $|\mathcal{X}_2|=4$ or $|\mathcal{X}_1|=8$, $|\mathcal{X}_2|=2$.
\end{remark}
We define $\mathcal{S}_X$ as all possible representations of $X$ where $X=(X_1,X_2)$. In other words we have $\mathcal{S}_X=\{(X_1,X_2):X=(X_1,X_2), \ |\mathcal{X}_1|\geq 2,\ |\mathcal{X}_2|\geq 2\}$.

Before stating the next theorem we derive an expression for $I(Y;U)$. We have
\begin{align}
I(Y;U)&=I(X,Y;U)-I(X;U|Y),\nonumber\\&=I(X;U)+I(Y;U|X)-I(X;U|Y),\nonumber\\&=I(X;U)\!+\!H(Y|X)\!-\!H(Y|U,X)\!-\!I(X;U|Y).\label{key}
\end{align}
As argued in \cite{kostala}, \eqref{key} is an important observation to find lower and upper bounds for $h_{\epsilon}(P_{XY})$. Next, we derive lower and upper bounds on $h_{\epsilon}(P_{XY})$. For deriving new lower bounds we use EFRL, ESFRL combining with separation technique.
\begin{theorem}\label{th.1}
	For any $0\leq \epsilon< I(X;Y)$ and pair of RVs $(X,Y)$ distributed according to $P_{XY}$ supported on alphabets $\mathcal{X}$ and $\mathcal{Y}$ we have
	\begin{align}\label{th2}
	\max\{L_1^{\epsilon},L_2^{\epsilon},L_3^{\epsilon},L_4^{\epsilon},L_5^{\epsilon}\}\leq h_{\epsilon}(P_{XY})\leq U_1^{\epsilon},
	\end{align}
	where
	\begin{align*}
	U_1^{\epsilon} &= H(Y|X)-\epsilon,\\
	L_1^{\epsilon} &= H(Y|X)-H(X|Y)+\epsilon=H(Y)-H(X)+\epsilon,\\
	L_2^{\epsilon} &= H(Y|X)-\alpha H(X|Y)+\epsilon\\&\ -(1-\alpha)\left( \log(I(X;Y)+1)+4 \right),\\
	L_3^{\epsilon} &= \epsilon\frac{H(Y)}{I(X;Y)}+g_0(P_{XY})\left(1-\frac{\epsilon}{I(X;Y)}\right),\\
	L_{4}^{\epsilon} &= H(Y|X)+\epsilon-\left( \log(I(X;Y)+1)+4 \right)\\&-\min_{(X_1,X_2)\in \mathcal{S}_X} \{\alpha_2 H(X_2|Y)\}, \\
	L_{5}^{\epsilon} &=H(Y|X)+\epsilon\\&-\!\!\!\!\!\!\!\!\min_{(X_1,X_2)\in \mathcal{S}_X}\!\!\!\{(1\!-\!\alpha_2)\!\left( \log(I(X;Y)\!+\!1)\!+\!4 \right)\!+\!\alpha_2 H(X|Y)\},
\end{align*}
and $\alpha=\frac{\epsilon}{H(X)}$, $\alpha_2=\frac{\epsilon}{H(X_2)}$ for any representation $X=(X_1,X_2)$. Furthermore, $g_{0}(P_{XY})=\max_{\begin{array}{c} 
	\substack{P_{U|Y,X}: I(U;X)=0,\\ X-Y-U}
	\end{array}}I(Y;U).$
The lower bound in \eqref{th2} is tight if $H(X|Y)=0$, i.e., $X$ is a deterministic function of $Y$. Furthermore, if the lower bound $L_1$ is tight then we have $H(X|Y)=0$. 
\end{theorem}
\begin{proof}
	The lower bound $L_3^{\epsilon}$ can be derived by using \cite[Remark~2]{shahab}, since we have $h_{\epsilon}(P_{XY})\geq g_{\epsilon}(P_{XY})\geq L_3^{\epsilon}$, where $g_{\epsilon}(P_{XY})=\sup_{\begin{array}{c} 
		\substack{P_{U|Y,X}: I(U;X)\leq\epsilon,\\ X-Y-U}
		\end{array}}I(Y;U)$. The upper bound $U_1^{\epsilon}$ and lower bounds $L_1^{\epsilon}$ and $L_2^{\epsilon}$ have been derived in \cite[Theorem 2]{king1}. The lower bound $L_1^{\epsilon}$ is attained by using \eqref{key} and EFRL. Similarly, the lower bound $L_2^{\epsilon}$ is attained by \eqref{key} and ESFRL, for more detail see \cite[Theorem 2]{king1}. Moreover, the results about tightness have been proved in \cite[Theorem 2]{king1}. It is sufficient to obtain $L_4^{\epsilon}$ and $L_5^{\epsilon}$. The complete proof for obtaining $L_4^{\epsilon}$ and $L_5^{\epsilon}$ is provided in Appendix A. As we mentioned earlier to achieve $L_1^{\epsilon}$ or $L_2^{\epsilon}$ we use EFRL or ESFRL. The main idea for constructing a RV $U$ that satisfies EFRL or ESFRL constraints is to add a randomized response to the output of FRL or SFRL. The randomization is taken over $X$. Now, let $(X_1,X_2)$ be a possible representation of $X$, i.e., $X=(X_1,X_2)$. The main idea to achieve $L_{4}^{\epsilon}$ and $L_{5}^{\epsilon}$ is to take randomization over $X_2$ instead of $X$. In other words we add a randomized response which is based on $X_2$ instead of $X$. Considering $L_2^{\epsilon}$ and $L_{4}^{\epsilon}$, $\alpha$ corresponds to the probability of randomizing over $X$, however, $\alpha_2$ corresponds to the probability of randomizing over $X_2$ for any representation $X=(X_1,X_2)$. 
	\end{proof}
In next corollary we let $\epsilon=0$ and derive lower bound on $h_0(P_{XY})$. 
\begin{remark}
	We emphasize that the lower bounds $L_1^{\epsilon}$, $L_2^{\epsilon}$, and $L_3^{\epsilon}$ have been derived in \cite{kostala}, \cite{king1}, and \cite{shahab}, respectively, however, the lower bounds $L_{4}^{\epsilon}$ and $L_{5}^{\epsilon}$ are obtained in this work. 
\end{remark}
\begin{remark}
	The lower bounds $L_1^{\epsilon}$, $L_2^{\epsilon}$, $L_4^{\epsilon}$, and $L_5^{\epsilon}$ have constructive proofs. Hence, statistical privacy mechanisms can be obtained using the lower bounds. For instance, RV $U$ that attains $L_4^{\epsilon}$ is built based on separation technique and extended version of SFRL. Noting that EFRL and ESFRL have also constructive proofs.
\end{remark}
\begin{corollary}(\cite[Corollary 2]{king1})
	If $X$ is a deterministic function of $Y$, the upper bound $U_1^{\epsilon}$ is attained. 
\end{corollary}
\begin{corollary}\label{kooni}
	Let $\epsilon=0$. In this case we obtain the same results as \cite[Corllary 1]{king1}, since $L_{4}^{0}=L_{5}^{0}=L_{2}^{0}$. For any pair of RVs $(X,Y)$ distributed according to $P_{XY}$ supported on alphabets $\mathcal{X}$ and $\mathcal{Y}$ we have
	\begin{align*}
	h_{0}(P_{XY})\geq \max\{L^0_1,L^0_2,g_{0}(P_{XY})\},
	\end{align*}
	where 
	\begin{align*}
	L^0_1 &= H(Y|X)-H(X|Y)=H(Y)-H(X),\\
	L^0_2 &= H(Y|X) -\left( \log(I(X;Y)+1)+4 \right).
	\end{align*}
\end{corollary}
Note that the lower bounds $L_{4}^{\epsilon}$ and $L_{5}^{\epsilon}$ do not lead to new bounds for $\epsilon=0$, however, for non-zero leakage they can improve the previous bounds. 
\section{comparison}\label{khaye}
In this part we study the bounds considering different cases. For simplicity let $X=(X_1,X_2)$ where $X_1$ and $X_2$ are arbitrary correlated. 
In this case we have
\begin{align*}
U_1^{\epsilon} &= H(Y|X_1,X_2)-\epsilon,\\
L_1^{\epsilon} &= H(Y|X_1,X_2)-H(X_1,X_2|Y)+\epsilon,\\
L_2^{\epsilon} &= H(Y|X_1,X_2)-\alpha H(X_1,X_2|Y)+\epsilon\\&\ -(1-\alpha)\left( \log(I(X_1,X_2;Y)+1)+4 \right),\\
L_3^{\epsilon} &= \epsilon\frac{H(Y)}{I(X_1,X_2;Y)}+g_0(P_{XY})\left(1-\frac{\epsilon}{I(X_1,X_2;Y)}\right),\\
L_{4}^{\epsilon} &= H(Y|X_1,X_2)+\epsilon-\left( \log(I(X_1,X_2;Y)+1)+4 \right)\\&-\alpha_2 H(X_2|Y), \\
L_{5}^{\epsilon} &=H(Y|X_1,X_2)+\epsilon\\&\!\!\!\!\!\!\!\!-(1\!-\!\alpha_2)\!\left( \log(I(X_1,X_2;\!Y)\!+\!1)\!+\!4 \right)\!+\!\alpha_2 H(X_1,X_2|Y),
\end{align*}
where $\alpha=\frac{\epsilon}{H(X)}$ and $\alpha_2=\frac{\epsilon}{H(X_2)}$.
We first compare the lower bounds $L_{1}^{\epsilon}$, $L_{4}^{\epsilon}$, and $L_{5}^{\epsilon}$. To do so, we consider four scenarios as follows.\\ 
\textbf{Scenario 1}: Let $X_1$ be a deterministic function of $Y$, i.e., $H(X_1|Y)=0$. Consequently, $L_{1}^{\epsilon}$ is dominant and we have $L_{1}^{\epsilon}\geq L_{5}^{\epsilon}\geq L_{4}^{\epsilon}$, since in this case $H(X_1,X_2|Y)=H(X_2|Y)$. \\
\textbf{Scenario 2}:
	Let $4+H(Y)\leq H(X_1|Y)$ and assume that $X_2$ is a deterministic function of $Y$, i.e., $H(X_2|Y)=0$. In this case, we have
	\begin{align}
	L_{4}^{\epsilon}\!-\!L_{5}^{\epsilon}\!&=\!\alpha_2\! \left(H(X_1|Y\!)\!-\!\log(I(X_1;Y\!)\!+\!H(X_2|X_1\!)\!+\!1)\!-\!4\right)\nonumber\\& \stackrel{(a)}{\geq} \alpha_2\left(H(X_1|Y)-I(X_1;Y)-H(X_2|X_1)-4\right)\nonumber\\&  \stackrel{(b)}{\geq} \alpha_2\left(H(X_1|Y)-I(X_1;Y)-H(Y|X_1)-4\right)\nonumber\\ &= \alpha_2\left( H(X_1|Y)-H(Y)-4\right)\nonumber\\&\geq 0,\label{jj}
	\end{align} 
	where (a) follows since $\log(1+x)\leq x$ and (b) holds since we have $H(X_2|X_1)\leq H(Y|X_1)$ and $H(X_2|Y)=0$. 
	Furthermore,
	\begin{align}
	L_{4}^{\epsilon}\!-\!L_{1}^{\epsilon}\nonumber&\!=\! H(X_1|Y)\!-\!\log(I(X_1;Y)\!+\!H(X_2|X_1)\!+\!1)\!-\!4\nonumber\\& \geq H(X_1|Y)-I(X_1;Y)-H(X_2|X_1)-4\nonumber\\& \geq H(X_1|Y)-I(X_1;Y)-H(Y|X_1)-4\nonumber\\ &=  H(X_1|Y)-H(Y)-4\nonumber\\&\geq 0.\label{jjj}
	\end{align}
	So, in this scenario $L_{4}^{\epsilon}$ is dominant and we have $L_{4}^{\epsilon}\geq \max(L_{1}^{\epsilon},L_{5}^{\epsilon})$. \\
\textbf{Scenario 3}:
Let $H(X_1,X_2)\geq 4$ and $Y$ be independent of $(X_1,X_2)$. In this case we have $L_{4}^{\epsilon}\geq L_{1}^{\epsilon}$ and $L_{5}^{\epsilon}\geq L_{1}^{\epsilon}$.\\
\textbf{Scenario 4}: Let $H(X_2|Y)\geq \log(I(X_2;Y)+1)+4$ and $X_1$ be a deterministic function of $X_2$. A simple example can be letting $X_1=f(X_2)$ and $H(X_2|Y)\geq \frac{H(X_2)}{2}+2$ which results in $H(X_2|Y)\geq \log(I(X_2;Y)+1)+4$ using $\log(x+1)\leq x$. We have
	\begin{align*}
	L_{5}^{\epsilon}-L_{4}^{\epsilon}&=\epsilon\frac{H(X_2|Y)}{H(X_2)}\\&+\epsilon \frac{\log(I(X_2;Y)+1)+4}{H(X_2)}-\epsilon\frac{H(X_2|Y)}{H(X_2)}\\ &= \epsilon \frac{\log(I(X_2;Y)+1)+4}{H(X_2)}\geq 0.
	\end{align*}
	Furthermore,
	\begin{align*}
	&L_{5}^{\epsilon}-L_{1}^{\epsilon}=\\&(1-\alpha)\left[H(X_2|Y)-\log(I(X_2;Y)+1)-4\right]\geq 0.
	\end{align*}
	Hence, in this case $L_{5}^{\epsilon}$ is dominant and we have $L_{5}^{\epsilon}\geq \max\{L_{4}^{\epsilon},L_{1}^{\epsilon}\}$. \\We next compare the lower bounds $L_{2}^{\epsilon}$, $L_{4}^{\epsilon}$, and $L_{5}^{\epsilon}$. To do so we consider the following scenarios.\\
	\textbf{Scenario 1}: To compare $L_{5}^{\epsilon}$ with $L_{2}^{\epsilon}$, let us assume that $H(X_1,X_2|Y)\leq \log(I(X_1,X_2;Y)+1)+4$. A simple example can be considering $X_1$ and $X_2$ as binary RVs. In this case we have
	\begin{align*}
	&L_{5}^{\epsilon}-L_{2}^{\epsilon}=\epsilon(\frac{1}{H(X_2)}-\frac{1}{H(X_1,X_2)})\times\\&\left(\log(I(X_1,X_2;Y)+1)+4-H(X_1,X_2|Y)\right)\geq 0.
	\end{align*}
	\textbf{Scenario 2}: To compare $L_{4}^{\epsilon}$ with $L_{2}^{\epsilon}$, let us assume that $X_2$ is a deterministic function of $Y$ and $H(X_1|Y)\geq \log(I(X_1,X_2;Y)+1)+4$. As we pointed out earlier a simple example is to let $4+H(Y)\leq H(X_1|Y)$ which leads to $H(X_1|Y)\geq \log(I(X_1,X_2;Y)+1)+4$. In this case we have
	\begin{align*}
	&L_{4}^{\epsilon}-L_{2}^{\epsilon}=\\&\frac{\epsilon}{H(X_1,X_2)}\left(H(X_1|Y)\!-\!\log(I(X_1,X_2;Y)+1)-4\right)\geq 0.
	\end{align*}
	Moreover, by using \eqref{jj} and \eqref{jjj} we conclude that $L_{4}^{\epsilon}$ is dominant and we have 
	\begin{align*}
	L_{4}^{\epsilon}\geq \max\{L_{2}^{\epsilon},L_{5}^{\epsilon},L_{1}^{\epsilon}\}.
	\end{align*} 
\section{conclusion}\label{concull}
We have introduced separation technique and it has been shown that combining it with extended versions of the FRL and SFRL, lead to new lower bounds on $h_{\epsilon}(P_{XY})$. 
If $X$ is a deterministic function of $Y$, then the bounds are tight. 
In different scenarios it has been shown that new bounds are dominant compared to the previous bounds.  
\bibliographystyle{IEEEtran}
\bibliography{IEEEabrv,IZS}
\newpage
\section*{Appendix A}
\subsection*{Deriving lower bounds $L_{4}^{\epsilon}$ and $L_{5}^{\epsilon}$:}
Let $(X_1,X_2)\in \mathcal{S}_X$, i.e., $(X_1,X_2)$ be a possible representation of $X$.
The bounds $L_{4}^{\epsilon}$ and $L_{5}^{\epsilon}$ can be obtained as follows. Let $\bar{U}$ be found by SFRL with $X=(X_1,X_2)$. 
We have
\begin{align*}
I(\bar{U};X_1,X_2)&=0,\\
H(Y|\bar{U},X_1,X_2)&=0,\\
I(X_1,X_2;\bar{U}|Y)&\leq \log(I(X_1,X_2;Y)+1)+4.
\end{align*}
Moreover, let $U=(\bar{U},W)$ with $W=\begin{cases}
X_2,\ \text{w.p}.\ \alpha_2\\
c,\ \ \text{w.p.}\ 1-\alpha_2
\end{cases}$, where $c$ is a constant which does not belong to $\mathcal{X}_1\cup \mathcal{X}_2 \cup \mathcal{Y}$ and $\alpha_2=\frac{\epsilon}{H(X_2)}$. First we show that $I(U;X_1,X_2)=\epsilon$. We have
\begin{align*}
I(U;X_1,X_2)&=I(\bar{U},W;X_1,X_2)\stackrel{(a)}{=}I(W;X_1,X_2)\\&=\!H\!(X_1,\!X_2)\!-\!\alpha_2 H(X_1|X_2)\!\\&-\!(1\!-\!\alpha_2)H\!(X_1,\!X_2)\\&=\alpha H(X_2)\\&=\epsilon,
\end{align*}
where (a) follows since $\bar{U}$ is independent of $(X_1,X_2,W)$.
Next, we expand $I(U;X_1,X_2|Y)$.
\begin{align}
&I(U;X_1,X_2|Y)\\&=I(\bar{U};X_1,X_2|Y)+I(W;X_1,X_2|Y,\bar{U})\nonumber\\&=I(\bar{U};X_1,X_2|Y)+H(X_1,X_2|Y,\bar{U})\!\\&-\!H(X_1,X_2|Y,\bar{U},W)\nonumber\\&=I(\bar{U};X_1,X_2|Y)+\alpha_2 H(X_1,X_2|Y,\bar{U})\\&-\alpha_2 H(X_1|Y,\bar{U},X_2)\nonumber \\&=I(\bar{U};X_1,X_2|Y)-\alpha_2 H(X_1|Y,\bar{U},X_2)\\&+\alpha_2\left( H(X_1,X_2|Y)-I(\bar{U};X_1,X_2|Y)\right)\nonumber\\&=\!(1\!-\!\alpha)I(\bar{U};X_1,\!X_2|Y)\!+\!\alpha_2 H(X_1,X_2|Y)\nonumber\\&-\!\alpha_2 H(X_1|Y\!,\!\bar{U}\!,\!X_2).\label{jakesh}
\end{align}
In the following we bound \eqref{jakesh} in two ways. We have
\begin{align}
\eqref{jakesh}&=\!(1\!-\!\alpha_2)I(\bar{U};X_1,\!X_2|Y)\!+\!\alpha_2 H(X_2|Y)\\&+\alpha I(X_1;\bar{U}|Y,X_2)\nonumber\\&=I(\bar{U};X_1,\!X_2|Y)\!+\!\alpha_2 H(X_2|Y)\!-\!\alpha_2 I(\bar{U};X_2|Y)\nonumber\\ &\stackrel{(a)}{\leq} \log(I(X_1,X_2;Y)+1)+4+\alpha_2 H(X_2|Y).\label{jakesh2}
\end{align}
Furthermore,
\begin{align}
\eqref{jakesh}&\leq \!(1\!-\!\alpha_2)I(\bar{U};X_1,\!X_2|Y)+\alpha_2 H(X_1,\!X_2|Y\!) \nonumber\\&\stackrel{(b)}{\leqq} \!(1\!-\!\alpha_2)\left(\log(I(X_1,X_2;Y)+1)+4\right)\nonumber\\&+\alpha_2 H(X_1,\!X_2|Y\!).\label{jakesh3}
\end{align}
Inequalities (a) and (b) follow since $\bar{U}$ is produced by SFRL, so that $I(\bar{U};X_1,X_2|Y)\leq \log(I(X_1,X_2;Y)+1)+4$. Using \eqref{jakesh2}, \eqref{jakesh3} and key equation in \eqref{key} we have
\begin{align}
h_{\epsilon}(P_{XY})&\geq I(U;Y)\nonumber\\&\stackrel{(c)}{\geq} \epsilon+H(Y|X_1,X_2)-\alpha_2 H(X_2|Y)\nonumber\\&-\left(\log(I(X_1,X_2;Y)+1)+4\right) \nonumber\\&=\epsilon+H(Y|X)-\alpha_2 H(X_2|Y)\nonumber\\&-\left(\log(I(X_1,X_2;Y)+1)+4\right),\label{as}
\end{align} 
and 
\begin{align}
h_{\epsilon}(P_{XY})&\geq I(U;Y)\nonumber\\&\stackrel{(d)}{\geq}  \epsilon+H(Y|X_1,X_2)-\alpha_2 H(X_1,X_2|Y)\nonumber\\&-(1-\alpha_2)(\log(I(X_1,X_2;Y)+1)+4)\nonumber \\&=\epsilon+H(Y|X)-\alpha_2 H(X|Y)\nonumber\\&-(1-\alpha_2)(\log(I(X;Y)+1)+4).\label{ass}
\end{align} 
In steps (c) and (d) we used $H(Y|X_1,X_2,U)=0$. The latter follows by definition of $W$ and the fact that $\bar{U}$ is produced by SFRL. Noting that since both \eqref{as} and \eqref{ass} hold for any representation of $X$ we can take maximum over all possible representations and we obtain
\begin{align*}
&h_{\epsilon}(P_{XY})\geq  H(Y|X)+\epsilon-\left( \log(I(X;Y)+1)+4 \right)\\&-\min_{(X_1,X_2)\in \mathcal{S}_X} \{\alpha_2 H(X_2|Y)\}=L_{4}^{\epsilon},\\
 &h_{\epsilon}(P_{XY})\geq H(Y|X)+\epsilon\\&-\!\!\!\!\!\!\!\!\min_{(X_1,X_2)\in \mathcal{S}_X}\!\!\!\{(1\!-\!\alpha_2)\!\left( \log(I(X;Y)\!+\!1)\!+\!4 \right)\!+\!\alpha_2 H(X|Y)\}\\&=L_{5}^{\epsilon}.
\end{align*}
\end{document}